\documentclass[twoside,12pt,fleqn,reqno,a4paper]{amsart}
\usepackage{amsmath}
\usepackage{epsfig} \usepackage{psfrag}\usepackage{subfigure}
\usepackage{tabularx}\usepackage{rotating}
\numberwithin{equation}{section}

\makeatletter
\renewcommand{\@thesubfigure}{(\alph{subfigure})}
\renewcommand{\p@subfigure}{}
\renewcommand{\@makecaption}[2]{%
\vskip 10\p@   \setbox\@tempboxa\hbox{#1.\space#2}%
\ifdim \wd\@tempboxa >\hsize       #1.\space#2\par     \else       \hbox to\hsize{\hfil\box\@tempboxa\hfil}%
\fi}
\def\citefull{\def\astroncite##1##2{##1 ##2}\@internalcite}\def\@cite#1#2{#1\if@tempswa  #2\fi}

\def\@fnsymbol#1{\ifcase#1\or \mbox{${^{\star}}$}\or
   \dagger\or \ddagger\or
   \S \or \P \or \|\or \mbox{$^{\star\star}$}\or \dagger\dagger
   \or \ddagger\ddagger\or \S\S\or \P\P\or \|\|\else ***
   \fi\relax}

\renewenvironment{equation}%
    {\@beginparpenalty\predisplaypenalty
     \@endparpenalty\postdisplaypenalty
     \refstepcounter{equation}%
     \trivlist \item[]\leavevmode
       \hb@xt@\linewidth\bgroup $\m@th
         \displaystyle }  
        {$\hfil 
         \displaywidth\linewidth\hbox{\@eqnnum}%
       \egroup
     \endtrivlist}

\makeatother
\newtheorem{theorem}{Theorem}[section]
\newtheorem{lemma}[theorem]{Lemma}
\theoremstyle{definition}

\theoremstyle{remark}

\numberwithin{equation}{section}

\begin{document}

\title[Relativistic Enskog Equation]
{Global Solution to the Relativistic Enskog Equation With Near-Vacuum Data}

\author{Zhenglu Jiang}
\address{Department of Mathematics, Zhongshan University,
Guangzhou 510275, P.~R.~China}
\email{mcsjzl@mail.sysu.edu.cn}

\subjclass[2000]{76P05; 35Q75; 82-02}

\date{November 7, 2006.}


\keywords{Relativistic; Enskog equation; global solution}

\begin{abstract}
We give two hypotheses of the relativistic collision kernal and show the existence and uniqueness of the global mild solution to
the relativistic Enskog equation with the initial data near the vacuum
for a hard sphere gas.
\end{abstract}

\maketitle
\section{Introduction}
\label{intro}
We will show the existence and uniqueness of the global mild solution to
the relativistic Enskog equation with near-vacuum initial data
for a hard sphere gas. By using a similar derivation
as for the relativistic Boltzmann equation [\cite{glw}],
the relativistic Enskog equation can be obtained as follows (see [\cite{gvo}]):
\begin{equation}
\frac {\partial f}{\partial t}+\frac {{\bf p}}
{p_0}\frac{\partial f}{\partial{\bf x}}
=Q(f)
\label{rbee}
\end{equation}
where $t\in [0,+\infty),$ ${\bf x}\in{\bf R}^3,$
${\bf p}\in{\bf R}^3,$ $p_0=(1+|{\bf p}|^2)^{1/2}$
and $Q(f)=Q(f)(t, {\bf x}, {\bf p})$ is the relativistic Enskog collision operator
which is expressed by the difference between the gain and loss terms
respectively defined by
 \begin{equation}
Q^+(f)(t,{\bf x},{\bf p})=\frac{a^2}{p_0}\int_{{\bf R}^3}\frac{d^3 {\bf p}_1}{p_{1 0}}
\int_{S_+^2}d\omega F^+(f)f(t,{\bf x},{\bf p}^\prime)f(t,{\bf x}+a{\bf \omega},{\bf p}_1^\prime)B(g, \theta)
\label{rbeekp}
\end{equation}
\begin{equation}
Q^-(f)(t,{\bf x},{\bf p})=\frac{a^2}{p_0}\int_{{\bf R}^3}\frac{d^3 {\bf p}_1}{p_{1 0}}
\int_{S_+^2}d\omega F^-(f)f(t,{\bf x},{\bf p})f(t,{\bf x}-a{\bf \omega},{\bf p}_1)B(g, \theta)
\label{rbeekm}
\end{equation}
where $ {f=f(t, {\bf x}, {\bf p})}$ is a distribution function of a
one-particle classical relativistic gas without external forces,
$a$ is a diameter of hard sphere ($a>0$) and
$F^\pm$ are two functionals on a Banach space $M$ defined in Section \ref{ao}.
More precisely speaking,
$F^\pm$ are high-density collision frequencies which
are physically defined by a geometrical factor $Y$ that
depends on the space density $\rho(t,{\bf x})=\int_{{\bf R}^3}f(t,{\bf x},{\bf p})d^3 {\bf p}$
at the time $t$ and the point of contact [e.g. $F^+=Y(\rho(t,{\bf x}+a\omega/2))$
and $F^-=Y(\rho(t,{\bf x}-a\omega/2))$].
The other different parts in equations (\ref{rbeekp})
and (\ref{rbeekm}) are explained below.

${\bf p}$ and ${\bf p}_1$ are dimensionless momenta
of two colliding particles immediately before collision while
${\bf p}^\prime$ and ${\bf p}_1^\prime$ are
 dimensionless  momenta  after collision;
$p_0=(1+|{\bf p}|^2)^{1/2}$ and $p_{10}=(1+|{\bf p}_1|^2)^{1/2}$ are
the dimensionless energy of two colliding particles imediately
before collision while $p_0^\prime=(1+|{\bf p}^\prime|^2)^{1/2}$ and
$p_{10}^\prime=(1+|{\bf p}_1^\prime|^2)^{1/2}$ are the dimensionless
energy after collision. ${\bf R}^3$ is a three-dimensional Euclidean
space and $S_+^2=\{\omega\in S^2 : \omega({\bf p}_1/p_{10}-{\bf
p}/p_0)\geq 0\}$ a subset of a unit sphere surface $S^2.$ $B(g,
\theta)$ is given by $B(g, \theta)= gs^\frac{1}{2}\sigma(g,
\theta)/2,$ where $\sigma(g, \theta)$ is a scattering cross section,
$s=|p_{10}+p_0|^2-|{\bf p}_1+{\bf p}|^2,$
$g=\sqrt{|{\bf p}_1-{\bf p}|^2-|p_{10}-p_0|^2}/2,$ $ \theta$ is the
scattering angle defined in $[0, \pi]$ by $\cos\theta
=1-2[(p_0-p_{10})(p_0-p_0^{\prime})-({\bf p}-{\bf p}_1)({\bf p}-{\bf
p}^{\prime})]/(4-s).$ Obviously, $s=4+4g^2$. ${\bf
\omega}=(\cos\psi\sin\theta,\sin\psi\sin\theta,\cos\theta)$ varies
on $S_+^2,$
 where $0{\leq}\theta{\leq}\pi,$ $ 0{\leq}\psi{\leq}2\pi. $
Put ${\bf p}^\prime={\bf p}+q\omega$ and $
{\bf p}_1^\prime={\bf p}-q\omega.$ Then,
by using a similar derivation as given by  Glassey and Strauss [\cite{gs95}]  for
the relativistic Boltzmann equation, we have
\begin{equation}
q\equiv q({\bf p},{\bf p}_1,\omega)
=\frac{2(p_0+p_{10})p_0p_{10}[\omega({\bf p}_1/p_{10}-{\bf p}/p_0)]
}{(p_0+p_{10})^2-[\omega({\bf p}+{\bf p}_1)]^2},\label{pq}
\end{equation}
\begin{equation}
g/s^\frac{1}{2}=8\frac{(p_0+p_{10})^2|\omega({\bf p}_1/p_{10}-{\bf p}/p_0)|
}{\{(p_0+p_{10})^2-[\omega({\bf p}+{\bf p}_1)]^2\}^2}.\label{peq}
\end{equation}
Here the scattering angle $\theta$ can be regarded as
a function of the variables ${\bf p},{\bf p}_1$ and $\omega,$ i.e., $\theta\equiv\theta({\bf p},{\bf p}_1,\omega).$
${\bf p}^\prime$ and ${\bf p}_1^\prime$ are bounded for bounded pre-collisional momenta
and lie on an ellipsoid when they are plotted in a plane, see detail explanation in [\cite{aci}].

The Moller velocity is defined as $v_M=gs^\frac{1}{2}/(p_0p_{10}),$
thus it can be found that $v_M^2=|{\bf p}/p_0-{\bf p}_1/p_{10}|^2-|{\bf p}\times{\bf p}_{1}/(p_0p_{10})|^2$ and that
$v_M\leq |{\bf p}/p_0-{\bf p}_1/p_{10}|.$

As a comparison the relativistic Boltzmann equation is the relativistic Enskog equation
with the factor $a^2F^\pm$ constant and the diameter $a$ equal to zero in the density variables.
Boltzmann's equation provides a successful description for dilute gases and
 is no longer valid when the density of the gas increases. The Enskog equation
proposed by Enskog [\cite{e22}] in 1922 is a modification of the Boltzmann equation
to explain the dynamical behavior of the density profile of a moderately dense gas.
It is thus a suitable idea that the relativistic Enskog equation is used to
model a hard sphere relativistic gas.

There are many results about
the relativistic Boltzmann equation, such as
global existence proof of Dudy\'{n}ski \& Ekiel-Je\.{z}ewska
[\cite{de88}, \cite{de92}] and properties
of the relativistic collision operator given by Glassey \& Strauss [\cite{gs91}],
 and background information of the classical Enskog equation
may be found in [\cite{a90}, \cite{e22}, \cite{p89}].
An existence and uniqueness theorem has been given
by Galeano et.~al~[\cite{gvo}] for the global solution to
the relativistic Enskog equation with data near the vacuum
for a hard sphere gas.
In their theorem (see Theorem 3.2 in [\cite{gvo}]),
a set $M_R$ is defined by $M_R= \{f\in M: ||f||\leq R\}$ with
\begin{equation}
R^2< \beta^4|v|/(16\pi^2cLa)
\label{gvoa01}
\end{equation}
and  an initial datum $f_0$ satisfies
\begin{equation}
||f_0||<Re^{\beta|x|^2}/2,
\label{gvoa02}
\end{equation}
where $M$ is defined by
$$M=\left\{f\in C([0,\infty)\times{\bf R}^3\times{\bf R}^3): \begin{array}{l}\hbox{ there exists }c>0 \hbox{ such that } \\
|f(t,x,v)|\leq c e^{-\beta(\sqrt{1+|v|^2}+|x+tv|^2)}\end{array}\right\}$$
with a norm
$$||f||=\sup\limits_{t,x,v}\{e^{\beta(\sqrt{1+|v|^2}+|x+tv|^2)}|f(t,x,v)|\},$$
$c,L,a$ and $\beta$ are positive constants, $t$ is a time variable in $[0,\infty),$
$x$ and $v$ are space and momentum variables in ${\bf R}^3$ respectively.
These assumptions imply that $M_R$ is an empty set and that none of the initial datum $f_0$ occurs.
Below let us prove this claim.
We first prove that $M_R$ is an empty set. Assume that $M_R$ is not an empty set.
Then there exists a function $f$ in $M_R$ such that $||f||\leq R.$ Thus $R\geq 0.$
By assumption (\ref{gvoa01}), we know that
\begin{equation}
R< \sqrt{\beta^4|v|/(16\pi^2cLa)}.
\label{gvoa03}
\end{equation}
Since $v$ is a momentum variables in ${\bf R}^3,$ by setting $v=0,$
(\ref{gvoa03}) shows that  $R<0$ as $v=0.$
This is in contradiction with $R\geq 0.$ Hence $M_R$ is an empty set.
Next, we show that none of the initial datum $f_0$ occurs.
Assume that there exists a function $f_0$ satisfying (\ref{gvoa02}).
It can be known from (\ref{gvoa02}) that $R> 0.$
By (\ref{gvoa01}), (\ref{gvoa03}) then follows.
Since $v$ is a momentum variables in ${\bf R}^3,$
by letting $v=0,$ (\ref{gvoa03}) shows that  $R<0$ as $v=0.$
This is in contradiction with $R> 0.$
None of the initial datum $f_0$ hence occurs.
If  (\ref{gvoa01}) and (\ref{gvoa02}) are replaced with
$R^2\leq \beta^4|v|/(16\pi^2cLa)$ and $||f_0||\leq Re^{\beta|x|^2}/2$ respectively,
then $M_R= \{0\}$ and  $f_0=0.$
Thus one can deduce that a unique solution to the relativistic Enskog equation
in their theorem is in fact zero. Hence this result is also trivial.
Notice that $v$  is a variable in ${\bf R}^3$ and that (\ref{gvoa01}) holds for all $v$ in ${\bf R}^3.$
If $|v|$ in (\ref{gvoa01}) is replaced with a positive constant $v_0,$ that is,
(\ref{gvoa01}) is changed as $R^2< \beta^4v_0/(16\pi^2cLa),$
then the problem is non-trivial.
Recently,  global existence of  mild solutions
has been proved by Glassey [\cite{g06}]  for
the relativistic Boltzmann equation with near-vacuum data
and  many relevant papers of both classical and relativistic cases
can be found in the reference.
Now there is not yet this result for the relativistic Enskog equation.
The aim of this paper is to extend this result into
the case of the relativistic Enskog equation.
In Section \ref{ao} two hypotheses of the relativistic collision kernel are given and
a Banach space and its operators are constructed. Then an existence and
uniqueness theorem of global mild solution to
the relativistic Enskog equation with near-vacuum data is given in Section \ref{eu}.

\section{Hypotheses and Operators}
\label{ao}
Let us begin by assuming that there is a positive function $m({\bf x},{\bf p})$
such that a nonnegative function $B(g,\theta)= gs^\frac{1}{2}\sigma(g, \theta)/2$ satisfies two hypotheses as follows:
\begin{eqnarray}
\frac{1}{p_0}\int_0^td\tau\int_{{\bf R}^3}\frac{d^3 {\bf p}_1}{p_{1 0}}
\int_{S_+^2}d\omega m({\bf x}+\tau{\bf p}/p_0-\tau{\bf p}^\prime/p_0^\prime,{\bf p}^\prime)\hspace*{4cm}
\nonumber \\
\times m({\bf x}+a{\bf \omega}+\tau{\bf p}/p_0-\tau{\bf p}_1^\prime/p_{10}^\prime,{\bf p}_1^\prime)
B(g, \theta)\leq m({\bf x},{\bf p})K,
\label{a01}
\end{eqnarray}
\begin{equation}
\frac{1}{p_0}\int_0^td\tau\int_{{\bf R}^3}\frac{d^3 {\bf p}_1}{p_{1 0}} \int_{S_+^2}d\omega
 m({\bf x}-a{\bf \omega}+\tau{\bf p}/p_0-\tau{\bf p}_1/p_{10},{\bf p}_1)B(g, \theta)\leq K.
\label{a02}
\end{equation}
for any ${\bf x}\in {\bf R}^3,$ ${\bf p}\in {\bf R}^3,$ $t\in{\bf R}^{+}$
and some positive constant $K.$
It can be known from the recent work of Glassey [\cite{g06}] that
there exist  two such functions $m({\bf x},{\bf p})$ and $B(g,\theta)$
satisfying (\ref{a01}) and (\ref{a02}).  For exampe, as given by Glassey [\cite{g06}],
we assume that
\begin{equation}
m({\bf x},{\bf p})=(1+|{\bf x}\times{\bf p}|)^{-(1+\delta)/2}e^{-p_0},
\label{a01ex}
\end{equation}
\begin{equation}
\sigma\equiv\sigma({\bf p},{\bf p}_1,\omega)=|\omega({\bf p}_1\times{\bf p})
|\tilde{\sigma}(\omega)/[p_{10}g(1+g^2)^{\delta+1/2}],
\label{a02ex}
\end{equation}
for any fixed $\delta\in (0,1),$ where $\tilde{\sigma}(\omega)$ is a nonnegative, bounded and continuous
function such that $\int_{S_+^2}\tilde{\sigma}(\omega)/(1+|{\bf z}\omega|)d\omega\leq c_0|{\bf z}|^{-1}$
for some positive constant $c_0$ and every non-zero element ${\bf z}\in {\bf R}^3.$
Thus a similar integral estimate to that developed by Glassey [\cite{g06}] leads to
the fact that (\ref{a01}) and (\ref{a02}) hold if $m({\bf x},{\bf p})$ and $B(g,\theta)$
are defined by (\ref{a01ex}) and (\ref{a02ex}) respectively. This indicates that our assumptions
(\ref{a01}) and (\ref{a02}) are valid for the relativistic Enskog equation.

Then we can construct a subset $M$ of a Banach space
$C([0,\infty)\times{\bf R}^3\times{\bf R}^3),$ which has the property that
every element $f=f(t,{\bf x},{\bf p})\in M$ if and only if there exists a positive constant $c$ such that $f$ satisfies
$|f^\#(t,{\bf x},{\bf p})|\leq c m({\bf x},{\bf p}),$  where and below everywhere,
 $f^\#$  is expressed as
$ f^\#(t, {\bf x}, {\bf p})=f(t, {\bf x}
+t{\bf p}/{p_0}, {\bf p})$ for any measurable function $f$ on
$(0, +\infty){\times}{\bf R}^3{\times}{\bf R}^3.$
It follows that $M$ is a Banach space when it has a norm
$||f||=\sup\limits_{t,{\bf x},{\bf p}}\{|f^\#(t,{\bf x},{\bf p})|m^{-1}({\bf x},{\bf p})\}.$
This space will be used below.

The relativistic Enskog equation (\ref{rbee})
can be also written as
$$\frac{d}{dt}f^\#(t, {\bf x}, {\bf p})=Q(f)^\#(t, {\bf x}, {\bf p}),$$
which leads to the following integral equation
\begin{equation}
f^\#(t, {\bf x}, {\bf p})=f_0({\bf x}, {\bf p})+\int_0^tQ(f)^\#(\tau, {\bf x}, {\bf p})d\tau.
\label{mild}
\end{equation}
A function $f(t, {\bf x}, {\bf p})$ is called global mild solution to the Enskog equation (\ref{rbee})
if $f(t, {\bf x}, {\bf p})$ satisfies the above integral equation (\ref{mild})
for almost every $(t, {\bf x}, {\bf p})\in [0,+\infty)\times{\bf R}^3\times{\bf R}^3.$
The definition of the term ``mild solution'' also appears
in the famous work of DiPerna and Lions [\cite{dl}]
where they show a global existence proof for the classical Boltzmann equation.

By (\ref{rbeekp}) and (\ref{rbeekm}), $Q(f)^\#(t, {\bf x}, {\bf p})$
can be rewritten as the difference between the gain and loss terms of  two other forms
 \begin{eqnarray}
Q^+(f)^\#(t,{\bf x},{\bf p})  =\frac{a^2}{p_0}\int_{{\bf R}^3}\frac{d^3 {\bf p}_1}{p_{1 0}}
\int_{S_+^2}d\omega F^+(f)f^\#(t,{\bf x}+t{\bf p}/p_0-t{\bf p}^\prime/p_0^\prime,{\bf p}^\prime)
\nonumber \\
\times f^\#(t,{\bf x}+a{\bf \omega}+t{\bf p}/p_0-t{\bf p}_1^\prime/p_{10}^\prime,{\bf p}_1^\prime)
B(g, \theta),
\label{rbeekph}
\end{eqnarray}
\begin{eqnarray}
Q^-(f)^\#(t,{\bf x},{\bf p}) \hspace*{0.7\linewidth}\nonumber \\
=\frac{a^2}{p_0}\int_{{\bf R}^3}\frac{d^3 {\bf p}_1}{p_{1 0}} \int_{S_+^2}d\omega
 F^-(f)f^\#(t,{\bf x},{\bf p})f^\#(t,{\bf x}-a{\bf \omega}+t{\bf p}/p_0-t{\bf p}_1/p_{10},{\bf p}_1)B(g, \theta).
\label{rbeekmh}
\end{eqnarray}

According to (\ref{rbeekph}) and (\ref{rbeekmh}),
we can finally build a Banach space $\widetilde{M}$ defined by $\widetilde{M}=\{f^\#:f\in M\}$ with
a norm $|||f^\#|||=||f||$ and
an operator $J$ on $\widetilde{M}$ by
\begin{equation}
J(f^\#)=f_0({\bf x}, {\bf p})+\int_0^tQ(f)^\#(\tau, {\bf x}, {\bf p})d\tau,
\label{op}
\end{equation}
since $F^\pm$ can be in fact regarded as two functionals on $\widetilde{M}.$

\section{Existence and Uniqueness}
\label{eu}
Let $M_R$ be denoted by $M_R=\{f\in M: ||f||\leq R\}$ for any $R\in {\bf R}_+,$
where $M$ is given in Section~\ref{ao}.
We first have the following lemma:
\begin{lemma}
Suppose that  the conditions (\ref{a01}) and (\ref{a02}) hold and that
$F^\pm$ are two functionals on $M_R$ such that
$|F^\pm(f)-F^\pm(g)|\leq L(R)||f-g||$ for any $f,g\in M_R$ where
$L(R)$ is a positive nondecreasing function on ${\bf R}_+.$
Then
$$\int_0^t|Q^+(f)^\#(\tau,{\bf x},{\bf p})|d\tau\leq C(R)m({\bf x},{\bf p})||f||^2,$$
$$\int_0^t|Q^-(f)^\#(\tau,{\bf x},{\bf p})|d\tau\leq C(R)m({\bf x},{\bf p})||f||^2$$
for any $f\in M_R,$ where $C(R)$ is a positive nondecreasing function on ${\bf R}_+.$
\label{lem}
\end{lemma}
\begin{proof}
It can be first found from the assumption of the two functionals $F^\pm$ that
there exists a positive constant $\tilde{L}(R)=L(R)R+|F^+(0)|+|F^-(0)|$
such that $|F^\pm(f)| \leq \tilde{L}(R)$ for any $f\in M_R.$
It follows from  (\ref{rbeekph}) and (\ref{rbeekmh}) that
 \begin{eqnarray}
\int_0^tQ^+(f)^\#(\tau,{\bf x},{\bf p}) d\tau
\leq\frac{\tilde{L}(R)a^2}{p_0}\int_0^td\tau\int_{{\bf R}^3}\frac{d^3 {\bf p}_1}{p_{1 0}}
\int_{S_+^2}d\omega ||f||^2\hspace*{3cm}
\nonumber \\
\times m({\bf x}+\tau{\bf p}/p_0-\tau{\bf p}^\prime/p_0^\prime,{\bf p}^\prime)
m({\bf x}+a{\bf \omega}+\tau{\bf p}/p_0-\tau{\bf p}_1^\prime/p_{10}^\prime,{\bf p}_1^\prime)
B(g, \theta),
\label{rbeekphintt}
\end{eqnarray}
\begin{eqnarray}
\int_0^tQ^-(f)^\#(\tau,{\bf x},{\bf p})d\tau
\leq\frac{\tilde{L}(R)a^2}{p_0}\int_0^td\tau\int_{{\bf R}^3}\frac{d^3 {\bf p}_1}{p_{1 0}} \int_{S_+^2}d\omega
 ||f||^2\hspace*{3cm}\nonumber \\
\times m({\bf x},{\bf p})m({\bf x}-a{\bf \omega}+\tau{\bf p}/p_0-\tau{\bf p}_1/p_{10},{\bf p}_1)B(g, \theta).
\label{rbeekmhintt}
\end{eqnarray}
By (\ref{a01}) and (\ref{a02}), (\ref{rbeekphintt}) and (\ref{rbeekmhintt}) give
\begin{eqnarray}
\int_0^tQ^+(f)^\#(\tau,{\bf x},{\bf p}) d\tau
\leq \tilde{L}(R)a^2Km({\bf x},{\bf p})||f||^2,
\nonumber
\end{eqnarray}
\begin{eqnarray}
\int_0^tQ^-(f)^\#(\tau,{\bf x},{\bf p})d\tau
\leq \tilde{L}(R)a^2Km({\bf x},{\bf p})||f||^2.
\nonumber
\end{eqnarray}
Take $C(R)=\tilde{L}(R)a^2K.$  It follows obviously that Lemma \ref{lem} holds.
\end{proof}

Then we can get the following theorem:
\begin{theorem}
Suppose that  the conditions (\ref{a01}) and (\ref{a02}) hold and that
$F^\pm$ are two functionals on $M_R$ such that
$|F^\pm(f)-F^\pm(g)|\leq L(R)||f-g||$ for any $f,g\in M_R$ where
$L(R)$ is a positive nondecreasing function on $R_+.$
 Then there exists a positive constant
$R_0$ such that the relativistic Enskog equation (\ref{rbee})
 has a unique  non-negative global mild solution $f=f(t,{\bf x},{\bf p})\in M_{R_0}$
through a non-negative initial data $f_0=f_0({\bf x},{\bf p})$ when
$\sup\limits_{{\bf x},{\bf p}}\{f_0({\bf x},{\bf p})m^{-1}({\bf x},{\bf p})\} $ is
 sufficiently small.
\label{th}
\end{theorem}
Theorem \ref{th} shows that there exists a unique  global mild solution to
the relativistic Enskog equation (\ref{rbee})
with the initial data near vacuum if a suitable assumption of
the scattering kernel is given.
Below is our proof of Theorem \ref{th}.
\begin{proof}
We first define a set $\widetilde{M}_R$ by
$\widetilde{M}_R=\{f^\#:|||f^\#|||\leq R, f^\#\in \widetilde{M}\},$
where $\widetilde{M}$ is given in Section~\ref{ao}.
By (\ref{op}) and Lemma~\ref{lem}, we have
\begin{equation}
|J(f^\#)|m^{-1}({\bf x},{\bf p})\leq |f_0({\bf x},{\bf p})
|m^{-1}({\bf x},{\bf p})+2C(R)||f||^2\leq R/2+2C(R)R^2
\nonumber
\end{equation}
for any $f^\#\in \widetilde{M}_R$ and $f_0$ with $||f_0||\leq R/2.$
Since $C(R)$ is a positive nondescreasing function on ${\bf R}_+,$ it follows that
$|||J(f^\#)|||\leq R$ for sufficiently small $R.$ Therefore $J$ is an operator from
$\widetilde{M}_R$ to itself for sufficiently small $R.$ Similarly, it can be also found that
$J$ is a contractive operator on $\widetilde{M}_R$ for some suitably small $R.$
Thus there exists a unique element $f^\#\in \widetilde{M}_R$ such that $f^\#=J(f^\#),$ i.e.,
(\ref{mild}) holds.
It then follows from the same argument as the one in [\cite{b67}]
(or see [\cite{g06}, \cite{is84}, \cite{ukai86}]) that
if $f_0({\bf x}, {\bf p})\geq 0$ then $f(t,{\bf x}, {\bf p})\geq 0.$
Hence the proof of Theorem \ref{th} is finished.
\end{proof}

\section*{Acknowledgement}
The author is supported by NSFC 10271121, 10511120278 and 10611120371,  and sponsored by
SRF for ROCS, SEM.
The author would like to thank the referees of this paper for their valuable comments on this work.



\begin{thebibliography}{100}
\bibitem[1]{aci}Andreasson H., Calogero S., Illner R.,
On blowup for gain-term-only classical and relativistic Boltzmann equations,
Math. Meth. Appl. Sci. 2004, {\bf 27}: 2231-2240.
\bibitem[2]{a90}Arkeryd, L., On the Enskog equation with large initial data,
SIAM Journal on Mathematical Analysis, 1990, {\bf 21}: 631 - 646
\bibitem[3]{b67}Bichteler K.,  On the Cauchy problem of the relativistic
Boltzmann equation,  Commun. math.~phys., 1967, {\bf 4}:352-364.
\bibitem[4]{glw}de Groot S.~R.,  Van Leeuwen W.~A.,  Van  Weert Ch.~G., Relativistic Kinetic
Theory, North-Holland, Amsterdam, 1980.
\bibitem[5]{dl}DiPerna R.~J., Lions P.~L., On the Cauchy problem for  Boltzmann
equations: Global existence and weak stability, Ann.~Math., 1989, {\bf 130}: 321.
\bibitem[6]{de88}Dudy\'{n}ski M., Ekiel-Je\.{z}ewska M.~L.,   On
 the  Linearized
Relativistic Boltzmann Equation, Commun. math. phys. 1988, {\bf 115}: 607-629.
\bibitem[7]{de92}Dudy\'{n}ski M., Ekiel-Je\.{z}ewska M.~L., Global Existence
 Proof  for Relativistic Boltzmann Equation,
Journal of Stat. Phys., 1992, {\bf 66} (3/4).
\bibitem[8]{e22}Enskog D., Kinetiche Theorie der W\`{a}rmeleitung,
Reibung und Selbstdiffusion in gewissen werdichteten
Gasen und Flubigkeiten, Kungl. Sv. Vetenskapsakademiens Handl. 63 (1922), 3-44,
English Transl. in Brush, S.~G., Kinetic Theory, vol 3, Pergamon, New York 1972.
\bibitem[9]{g06}Glassey R.,  Global solutions to the Cauchy problem for the relativistic
Boltzmann equation with near-vacuum data, Comm.~Math.~Phys., 2006, {\bf 26}: 705-724.
\bibitem[10]{gs91}Glassey R., Strauss W., On the derivatives of the collision  map
of relativistic particles, Transp.~Theory Stat.~Phys., 1991, {\bf 20}: 55-68.
\bibitem[11]{gs95}Glassey R., Strauss W., Asymptotic stability of the relativistic Maxwellian
via fourteen moments, Transp.~Theory Stat.~Phys., 1995, {\bf 24}: 657-678.
\bibitem[12]{gvo}Galeano R., Vasquez O., Orozco B.,
the relativistic Enskog equation, Journal of Differential Equations, Conference 13, 2005, pp 21-27.
\bibitem[13]{is84}Illner, R., Shinbrot, M., The Boltzmann Equation, global existence
for a rare gas in an infinite vacuum, Comm.~Math.~Phys., 1984, {\bf 95}: 217-226.
\bibitem[14]{p89}Polewczak, J., Global existence and asymptotic behavior for the nonlinear Enskog equation,
SIAM Journal on Applied Mathematics, 1989, {\bf 49}: 952 - 959
\bibitem[15]{ukai86}Ukai, S., Solutions of the Boltzmann Equation, Studies in Math.~Appl., 1986, {\bf 18}: 37-96.
\end{thebibliography}
\end{document}